\documentclass[]{aa}  

\usepackage{graphicx}
\usepackage{txfonts}
\usepackage{natbib}
\defcitealias{1990ApJ...358..328G}{GF90}
\defcitealias{2012A&A...539A..39C}{CL12}
\defcitealias{2022A&A...661A..77H}{HCD22}

\usepackage{ulem}
\usepackage{multirow}

\graphicspath{{./}{figures/}}

\usepackage[colorlinks, citecolor=blue]{hyperref}
\begin{document}

   \title{Evaluation of different recipes for chromospheric radiative losses in solar flares}
   
   \author{J. Tian\inst{1,2}
          \and
          J. Hong\inst{3,4}
          \and
          Y. Li \inst{1,2}
          \and
          M. D. Ding \inst{3,4}
          }

   \institute{Key Laboratory of Dark Matter and Space Astronomy, Purple Mountain Observatory, CAS, Nanjing 210023, PR China
   \and
   School of Astronomy and Space Science, University of Science and Technology of China, Hefei 230026, PR China
   \and
   School of Astronomy and Space Science, Nanjing University, Nanjing 210023, PR China\\
   \email{jiehong@nju.edu.cn}
   \and
   Key Laboratory for Modern Astronomy and Astrophysics (Nanjing University), Ministry of Education, Nanjing 210023, PR China
             }


 
  \abstract
   {Radiative losses are an indispensable part in the numerical simulation of flares. Detailed calculations could be computationally expensive, especially in the chromosphere. There have been some approximate recipes for chromospheric radiative losses in flares, yet their feasibility in flare simulations needs further evaluation.}
   {We aim to evaluate the performance of different recipes for chromospheric radiative losses in flare simulations.}
   {We compare the atmospheric structure and line profiles in beam-heated flares calculated with detailed radiative losses and the approximate recipes.}
   {Both \citetalias{1990ApJ...358..328G} and \citetalias{2022A&A...661A..77H} recipes provide acceptable total radiative losses compared with detailed one, but there are discrepancies in the different atmospheric layers during the different evolutionary phases, which leads to misestimations of temperature and line intensity. The recipe of \citetalias{1990ApJ...358..328G} overestimates the coolings in the upper chromosphere greatly when temperature exceeds 10$^{5}$ K, which also affects the flare evolution and line asymmetries. Radiative heating in the middle chromosphere only functions in the initial stage and could be safely neglected. However, radiative heating from Lyman continuum could dominate near the transition region.}
   {}


   \maketitle
%

\section{Introduction}
The interaction between radiation and matter is very important since radiation plays a vital role in the energy transport of the solar  atmosphere. Extinctions or emissions of photons from transitions between atomic energy levels can either heat or cool the local atmosphere, and their contributions are expressed as the radiative flux divergence, referred to as the radiative losses.  In the photosphere, the assumption of local thermodynamic equilibrium (LTE) is a good approximation. The source function is equal to the Planck function and the opacity can be solved using the opacity distribution functions \citep{1966AJ.....71S.181S,1967ApJ...149..169M} or the multigroup method \citep{1982A&A...107....1N,1994A&A...284..105L,2000ApJ...536..465S}. Thus, the radiative losses can be calculated in a straightforward way following its definition. In the transition region and corona, the coronal approximation is accurate. A feasible way to calculate the radiative loss is to use the product of the electron density, the hydrogen density and the optically thin radiative loss function  \citep{2018LRSP...15....5D}. However, both assumptions break down in the chromosphere. To calculate the radiative losses in the chromosphere which is dominated by non-local thermodynamic equilibrium (non-LTE), the radiative transfer equation coupled to the population rate equation needs to be solved accurately, as in \cite{1981ApJS...45..635V}. 

Up to now, there are few numerical codes that could fully tackle the coupling of radiation and matter. In the 1D regime, the radiative hydrodynamic codes \verb"RADYN" \citep{1992ApJ...397L..59C,1995ApJ...440L..29C,1997ApJ...481..500C,2002ApJ...572..626C} and \verb"FLARIX" \citep{2009A&A...499..923K} can reproduce the evolution of the solar atmosphere in response to flare energy deposition, which can automatically calculate precise radiative losses from non-LTE solutions of the radiative transfer equation. However, such approaches become quite computationally expensive in the 2D and 3D regimes \citep{2008PhST..133a4012C}. Fortunately, there have been efforts in the past decades aimed to construct a simple recipe for chromospheric radiative losses.

\citet[][hereafter \citetalias{1990ApJ...358..328G}]{1990ApJ...358..328G} improved the recipe of \citet{1980SoPh...68..351N} in the calculations of radiative cooling based on semi-empirical flare models. It has been employed in the numerical simulations of flares \citep{1991A&A...241..618G,1994Ap&SS.213..233D,2010ApJ...710.1387J} as well as other activities such as Ellerman bombs and ultraviolet bursts \citep{2001ChJAA...1..176C,2011RAA....11..225X,2015ApJ...799...79N,2016ApJ...832..195N,2021A&A...646A..88N} in the solar atmosphere. In addition, it is also used to estimate the radiative energy of chromospheric activities in semi-empirical modeling \citep{2006ApJ...643.1325F,2010RAA....10...83F,2017RAA....17...31F,2015RAA....15.1513L}. \citet[][hereafter \citetalias{2012A&A...539A..39C}]{2012A&A...539A..39C} proposed a  new formula where the chromospheric radiative losses are related to the optically thin emission, the escape probability and the ionization fraction. These three parameters are empirically tabulated from detailed non-LTE calculations. This approximation has been introduced into the \verb"Bifrost" code \citep{2011A&A...531A.154G}, the \verb"HYDRAD" code \citep{2013ApJ...770...12B}, the extended version of the \verb"MURaM" code \citep{2021A&A...656L...7C,2022arXiv220403126P} and the \verb"RAMENS" code \citep{2021ApJ...916L..10W} to compute the non-LTE radiative losses in the chromosphere. Recently, \citet[][hereafter \citetalias{2022A&A...661A..77H}]{2022A&A...661A..77H} updated the formula following  \citetalias{2012A&A...539A..39C} for calculating radiative losses in a flaring atmosphere.

As already mentioned by \citet{1990ApJ...358..328G}, \citetalias{1990ApJ...358..328G} underestimates radiative cooling in the lower chromosphere compared with detailed non-LTE calculations. \citet{2022A&A...661A..77H} also found that the \citetalias{1990ApJ...358..328G} cooling is weaker in the middle chromosphere, but stronger in the upper chromosphere. The recipe of \citetalias{1990ApJ...358..328G} is constructed from two semi-empirical flare models, and
the recipe of \citetalias{2022A&A...661A..77H} is constructed from flare models heated by non-thermal electrons with peak fluxes in the order of $\mathrm{10^{10}}$ $\mathrm{erg \cdot cm^{-2} \cdot s^{-1}}$. Since the flare conditions vary from case to case, the validity of these recipes remains to be verified. Although \citet{2022A&A...661A..77H} has briefly compared the performance of these recipes, they only calculated the radiative losses from the atmospheres that are pre-calculated with \verb"RADYN". A better evaluation would require inclusion of these recipes in real flare simulations.

In this paper, we investigate the properties of the lower atmosphere calculated using different recipes of radiative losses,  as well as the resulted line profiles, in order to explore the application range of these recipes. In Section \ref{Simulations}, we introduce our method. We compare models calculated with different recipes in Section \ref{compare}. A discussion of the results is followed in Section \ref{discussion} and finally the conclusions are in Section \ref{conclusion}.
\section{Method \label{Simulations}}
The radiative hydrodynamic code \verb"RADYN" is developed for analysis of chromospheric shocks at first \citep{1992ApJ...397L..59C,1995ApJ...440L..29C,1997ApJ...481..500C}, and later widely applied to flare simulations \citep{1999ApJ...521..906A,2005ApJ...630..573A,2015ApJ...809..104A}. 
Important transitions between energy levels for the most important atoms, including those from a six-level with continuum H atom, a six-level with continuum \ion{Ca}{II} atom and a nine-level with continuum He atom, are added in the condition of non-LTE. In the equation of internal energy conservation, the total optically thick radiative loss is calculated by summing up losses contributed by all the bound-bound and bound-free transitions. 

We modify the calculations of optically thick radiative loss in  \verb"RADYN"  by replacing it with the approximate recipes of \citetalias{1990ApJ...358..328G} and \citetalias{2022A&A...661A..77H}. For the recipe of \citetalias{1990ApJ...358..328G}, it is used to calculate the total optically thick radiative losses. For the recipe of \citetalias{2022A&A...661A..77H}, it is used to calculate only the losses contributed from lines and the Lyman continuum (LyC), while the contributions from other continua  are still from \verb"RADYN" itself. We note that \verb"RADYN" includes the He lines while the recipe of \citetalias{2022A&A...661A..77H} includes the Mg lines, and the implications of this difference are discussed in Section~\ref{4.1}. Moreover, we only modify chromospheric losses from 0.5 Mm to 1.8 Mm. However, the radiative heating in the upper chromosphere from LyC cannot be neglected at a certain time in the flare simulations, and we choose to retain this part. Possible influences are discussed further in Section~\ref{4.2}.

We rerun the flare models with different recipes for chromospheric radiative losses and compare the atmospheric evolutions and line profiles. Flare models are labeled in the form of FXn, where the letter X is from A to D for different heating parameters, which are summarized in Table \ref{tab:flare models}. The number n in the labels ranges from 0 to 2 for different treatments of the radiative losses, with 0 for the detailed treatment, 1 for the \citetalias{1990ApJ...358..328G} recipe, and 2 for the \citetalias{2022A&A...661A..77H} recipe. 
The heating parameters are specifically chosen for typical solar flares and to compare different recipes. Among all the heating parameters, the electron energy flux has the largest influence on flare evolution. Previous studies showed that the variation of electron energy flux could lead to a difference between gentle and explosive chromospheric evaporation \citet[e.g.][]{1985ApJ...289..434F,1985ApJ...289..425F,1985ApJ...289..414F}. Thus, we mainly vary the energy flux of electrons in our simulations. Besides, in the post-evaluation of radiative losses in \citet{2022A&A...661A..77H}, the performance of \citetalias{2022A&A...661A..77H} is the best in flare model FA0, while the performance of \citetalias{1990ApJ...358..328G} peaks in flare model FD0. These models are included in our simulations here in order to see if their behaviors would be different.
The energy flux of non-thermal electron follows a linearly increasing function over time for a period of 10.0 s and the other simulation setups are the same as \citet{2022A&A...661A..77H}. 

\begin{table*}[h]
    \caption{List of parameters of the flare models for simulations}
    \centering 
    \begin{tabular}{c c c c c} 
        \hline\hline  
        Label & Peak electron flux ($\mathrm{erg \cdot cm^{-2} \cdot s^{-1}}$) &Total duration (s)  & Spectral index & Cutoff energy (keV) 
        \\
        \hline  
        FA0--FA2 & $\mathrm{1 \times 10^{10}}$ & 10.0 & 3 & 25 \\
                
        FB0--FB2 & $\mathrm{1 \times 10^{11}}$ & 10.0 & 3 & 25 \\
                
        FC0--FC2 & $\mathrm{1 \times 10^{9}}$ & 10.0 & 3 & 5 \\
        
        FD0--FD2 & $\mathrm{1 \times 10^{10}}$ & 10.0 & 7 & 25 \\
        \hline 
    \end{tabular}
    \label{tab:flare models}
\end{table*}

\section{Results \label{compare}}
\subsection{Comparison of GF90 and RADYN}
\subsubsection{Atmospheric structure \label{gf90_as}}
\begin{figure*}
\includegraphics[width=18.4cm]{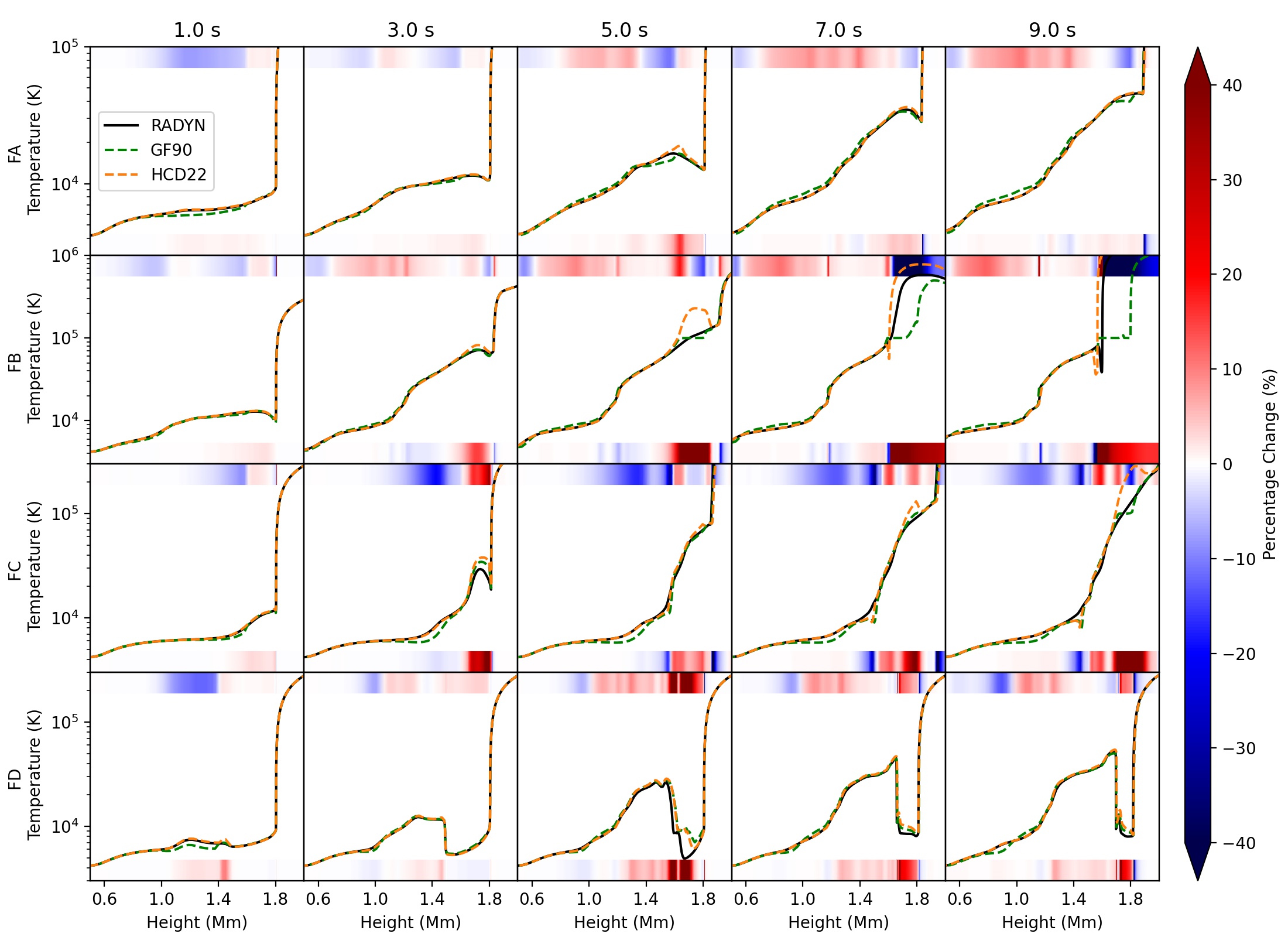}
\caption{Comparison of the evolution of the atmospheric structure. The top row shows the atmospheric temperature calculated with detailed treatment of the radiative processes by RADYN (black solid lines) as well as that calculated using the recipe of \citetalias{1990ApJ...358..328G} (green dashed lines) and \citetalias{2022A&A...661A..77H} (orange dashed lines) for Case FA. The following three rows are the same as the top row, but for Cases FB, FC and FD, respectively. The temperature deviation between \citetalias{1990ApJ...358..328G} and RADYN results is also shown as a horizontal bar at the top of each panel, and that between \citetalias{2022A&A...661A..77H} and RADYN results is shown at the bottom of each panel in a color scale.}
\label{fig:atmosphere}
\end{figure*}

\begin{figure*}
\includegraphics[width=18.4cm]{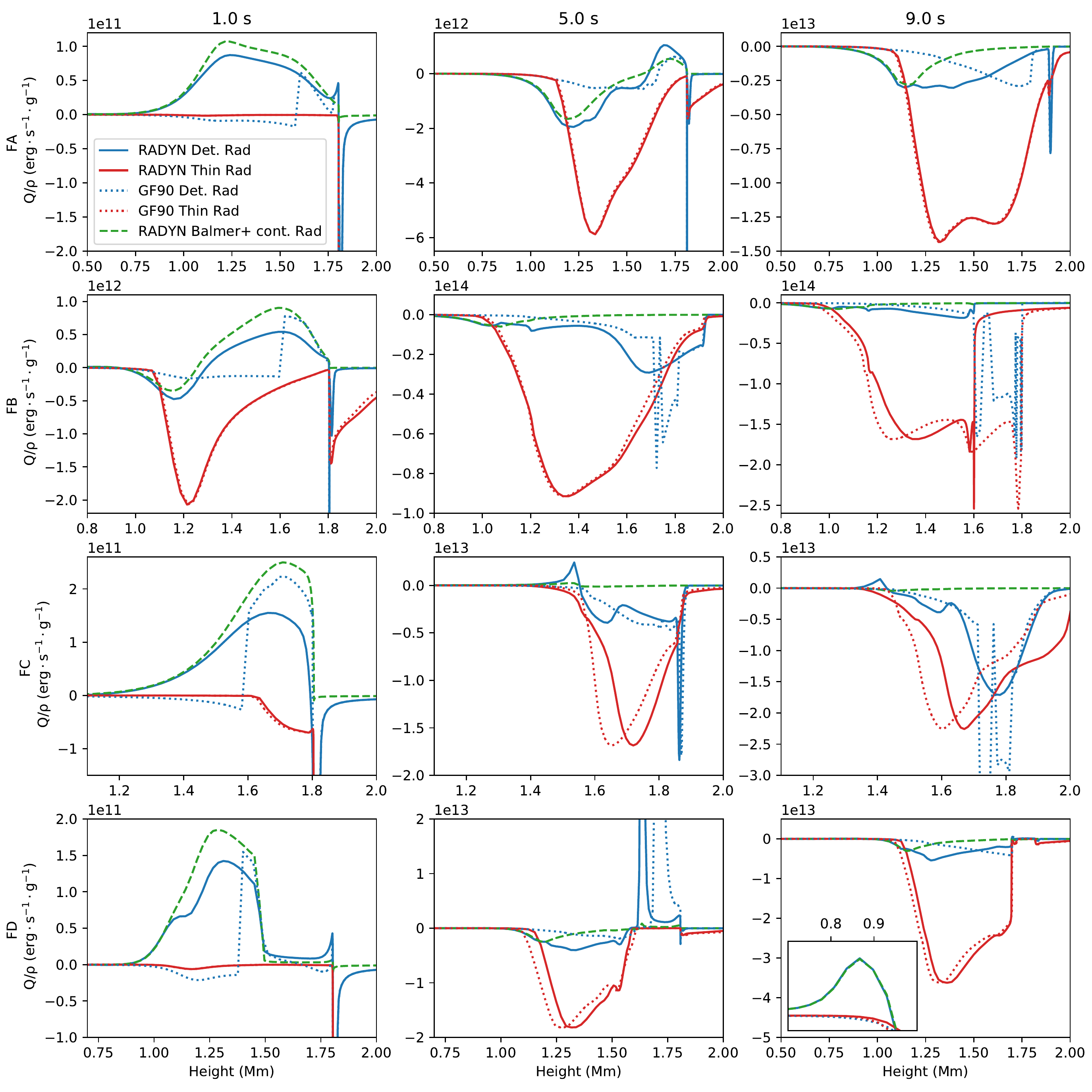}
\caption{Height distributions of the radiative losses per unit mass ($Q/\rho$) calculated with detailed treatment of the radiative processes by RADYN (solid lines) and by the recipe of \citetalias{1990ApJ...358..328G} (dotted lines). From top to bottom, the four rows represent the results for Cases FA, FB, FC and FD. The blue and red lines represent the losses calculated with detailed treatment and with optically thin assumption in each panel respectively. Also plotted is the radiative losses from H Balmer and higher continua (green dashed lines). 
A positive value means radiative heating, while a negative one means radiative cooling.}
\label{fig:radloss_gf90}
\end{figure*}

Figs. \ref{fig:atmosphere} and  \ref{fig:radloss_gf90} show the atmospheric structure and radiative losses for the four flare models calculated with detailed treatment (FA0--FD0) and the approximated recipe of \citetalias{1990ApJ...358..328G} (FA1--FD1) for radiative losses. The integrated chromospheric radiative losses are summarized in Table \ref{tab:tot_loss}. For the atmosphere calculated from detailed radiative losses (FA0--FD0), during the initial stage of a flare, $t<1.0$ s, non-thermal electron beams heat the region quickly, and the chromospheric temperature begins to rise. The electron number density increases dramatically, and at this stage, more than 90\% of the internal energy is stored as ionization energy. There are strong radiative heatings ($\sim1$ erg s$^{-1}$ g$^{-1}$) mainly contributed by Balmer and higher continua in the middle and upper chromosphere (1.0 Mm above, see the first column in Fig. \ref{fig:radloss_gf90}). For the  FA1--FD1 cases, the amount of ionization energy is the same as in the FA0--FD0 cases, as judged from the electron density in Fig. \ref{fig:density}, while the overestimation of radiative losses leads to an underestimated thermal energy. In consequence, the chromospheric temperature rises more slowly.

\begin{figure*}
\includegraphics[width=18.4cm]{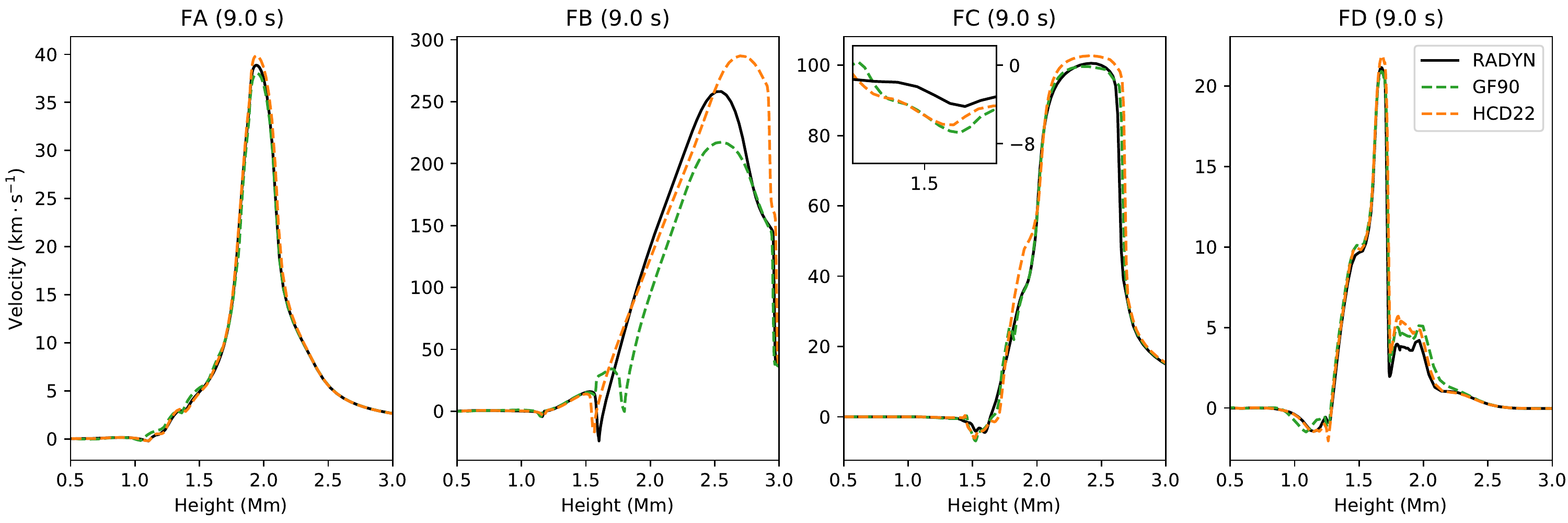}
\caption{Height distributions of velocity for four cases at 9.0 s. Positive velocity values denote upward motions, while negative velocity values denote downward motions. The black, green and orange lines represent the results of RADYN, \citetalias{1990ApJ...358..328G} and \citetalias{2022A&A...661A..77H} respectively.}
\label{fig:velocity}
\end{figure*}

As the heating continues, the chromospheric temperature and electron density increase persistently until most of the flare heating energy is lost through radiative cooling for all cases.  After 5.0 s, optically thin radiative cooling is dominating in upper chromosphere (above 1.2 Mm in Case FA) due to high temperature and increased electron density. We also notice that cooling from optically thin Balmer and higher continua exceeds cooling from optically thick lines in middle chromosphere (below 1.2 Mm in Case FA0) (see the middle and right columns in Fig. \ref{fig:radloss_gf90}). In Case FA1, chromospheric radiative cooling is underestimated in the height range of 0.7--1.6 Mm. As a result, the overestimated net energy input leads to a rise in temperature by about 10\%--20\%, as well as a rise in the electron number density by about 20\%--75\% 
as shown in Fig. \ref{fig:density}. While in the upper chromosphere (1.6--1.8 Mm), the optically thick radiative cooling is overestimated, making the region cooler. The increase in temperature forces chromospheric plasma to move upward into the corona, known as the chromospheric evaporation. Compared with Case FA0, Case FA1 generates a velocity of chromospheric evaporation about 1 km s$^{-1}$ smaller at 9.0 s as shown in Fig. \ref{fig:velocity}.

\begin{table}[h]
    \caption{Temporally and spatially integrated radiative losses (in units of 10$^9$ $\mathrm{erg \cdot cm ^{-2}}$) calculated with detailed treatment in different models. The integration height range is 0.5–1.8 Mm.}
    \centering 
    \begin{tabular}{c c c c c} 
        \hline\hline  
        & RADYN & GF90 & HCD22
        \\
        \hline  
        FA & 14.36 & 3.19 & 13.82 \\
                
        FB & 127.90 & 25.62 & 134.42 \\
                
        FC & 0.17 & 0.51 & -0.11 \\
        
        FD & 9.32 & 2.47 & 6.53 \\
        \hline 
    \end{tabular}
    \label{tab:tot_loss}
\end{table}

In Case FB0 heated by a larger peak electron flux, the upper chromosphere above 1.6 Mm is heated beyond $\mathrm{10^5}$ K after 5.0 s, resulting in an explosive chromospheric evaporation with a maximum velocity of about 260 km s$^{-1}$ and an accompanying chromospheric condensation with a maximum downward velocity of about 24 km s$^{-1}$. Near 1.6 Mm, there appears a high density, low temperature region known as the condensation region. In Case FB1, the upper chromosphere in the height range of 1.6--1.8 Mm is less heated, and the temperature does not reach $10^6$ K at 9.0 s, due to an overestimation of the optically thick cooling. Chromospheric evaporation with maximum velocity of about 220 km s$^{-1}$ at 9.0 s carries less materials into corona, and there appears to be no chromospheric condensation region, although there is a region near 1.8 Mm with a smaller velocity (about $-$0.2 km s$^{-1}$) than adjacent regions. In the lower and middle chromosphere (0.5--1.6 Mm), Case FB1 follows the pattern of FA1, with a warmer atmosphere and a higher electron number density compared with the FB0 and FA0 cases. 

For Case FC0, flare heating is centered on the upper chromosphere, and interestingly, there is a peak of radiative heating contributed by LyC near 1.5 Mm at 5.0 s.
We find a temperature dip in Case FC1 as a result of the absence of radiative heating in the approximated \citetalias{1990ApJ...358..328G} recipe. In addition, there is a region of high density near 1.53 Mm in Case FC1 accompanied with a larger downflow velocity of about 7 km s$^{-1}$ at 9.0 s compared with the 4 km s$^{-1}$ in Case FC0. The behavior of the upper chromosphere above 1.7 Mm is somewhat similar to other cases. 

In Case FD0, the heating is more concentrated and the heating range is smaller than that in Case FA0. Thus, the lower chromosphere below 0.9 Mm is nearly undisturbed where Balmer and higher continua still act as a radiative heating source as in the first 1.0 s of Case FA0. By comparison, the lower chromosphere in Case FD1 turns out to be cooler due to the absence of radiative heating. However, the temperature in the chromosphere above 1.2 Mm is higher for Case FD1, which is caused by the underestimation of the coolings in 1.2--1.6 Mm. As for the region above 1.6 Mm, the stronger heating from LyC brings a warmer atmosphere for Case FD1.

\subsubsection{Line profiles}
\begin{figure*}
\includegraphics[width=17cm]{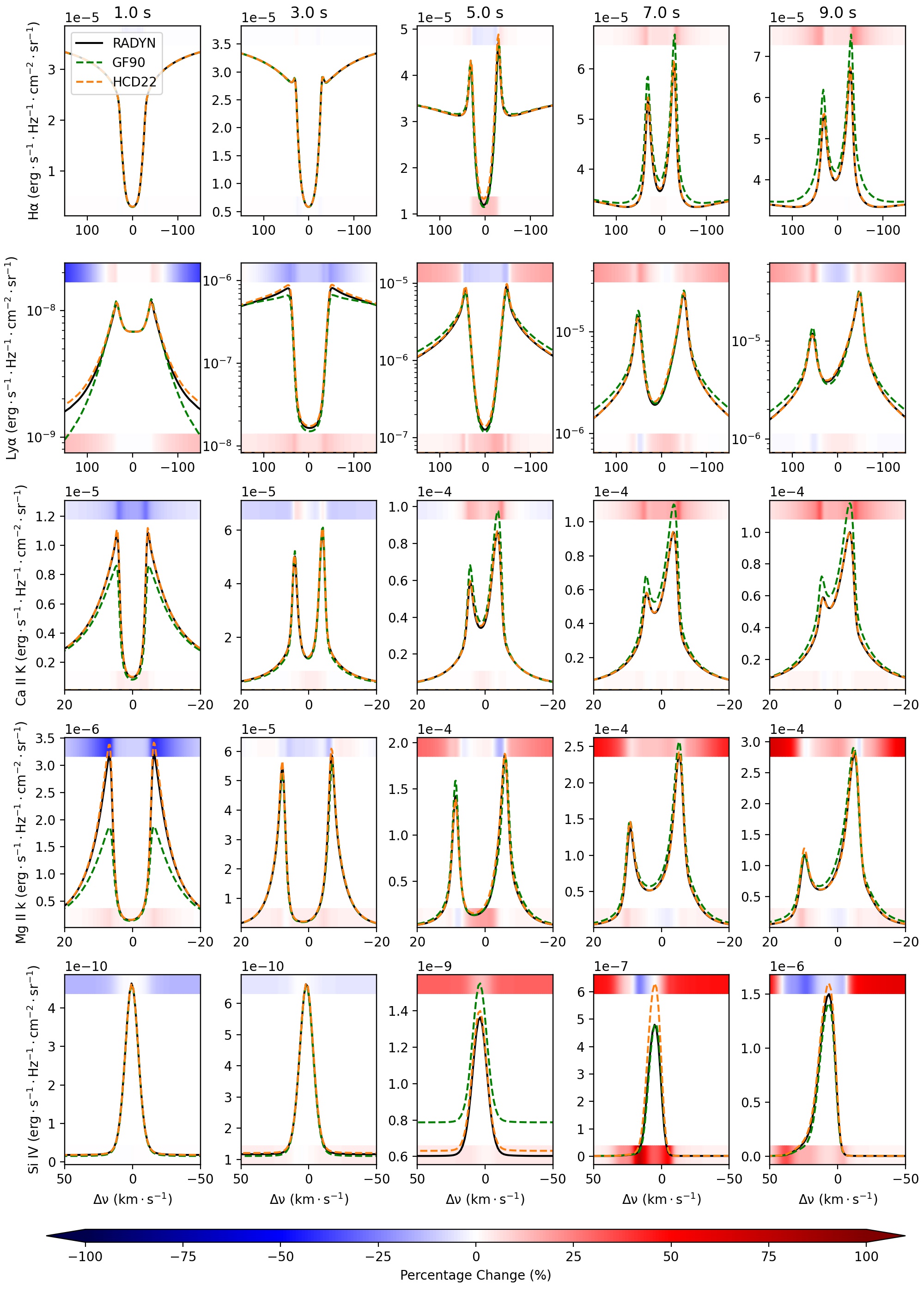}
\caption{Time evolution of the synthetic H$\alpha$, Ly$\alpha$, \ion{Ca}{II} K, \ion{Mg}{II} k and \ion{Si}{IV} line profiles for FA. The horizontal axes are in the Doppler scale. The deviation between the \citetalias{1990ApJ...358..328G} and RADYN solutions is also shown as a horizontal bar in the top part of each panel, while that between the \citetalias{2022A&A...661A..77H} and RADYN solutions is shown in the bottom part of each panel.}
\label{fig:spectram_fa}
\end{figure*}

\begin{figure*}
\includegraphics[width=17cm]{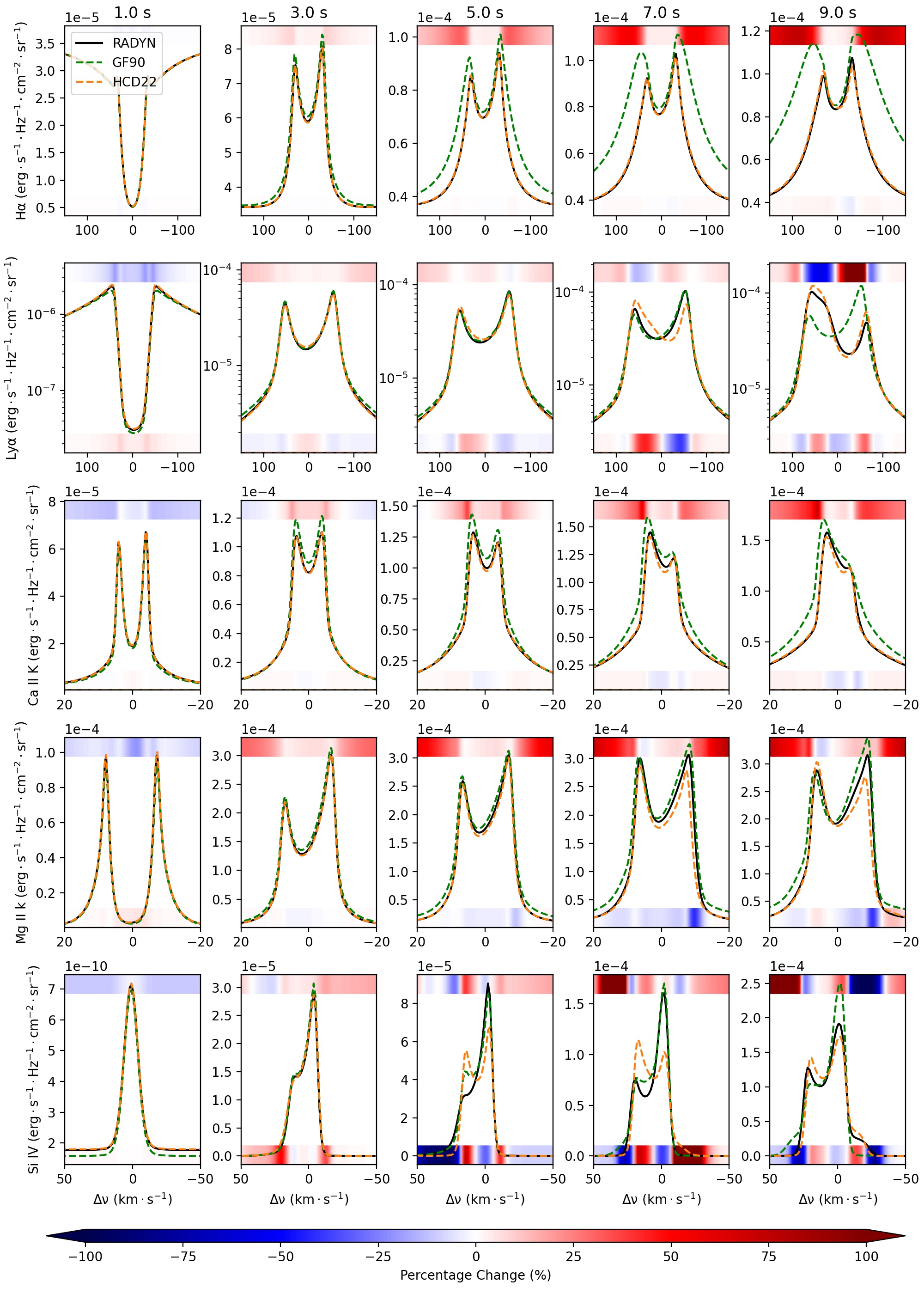}
\caption{Same as Figure \ref{fig:spectram_fa}, but for FB.}
\label{fig:spectram_fb}
\end{figure*}

\begin{figure*}
\includegraphics[width=17cm]{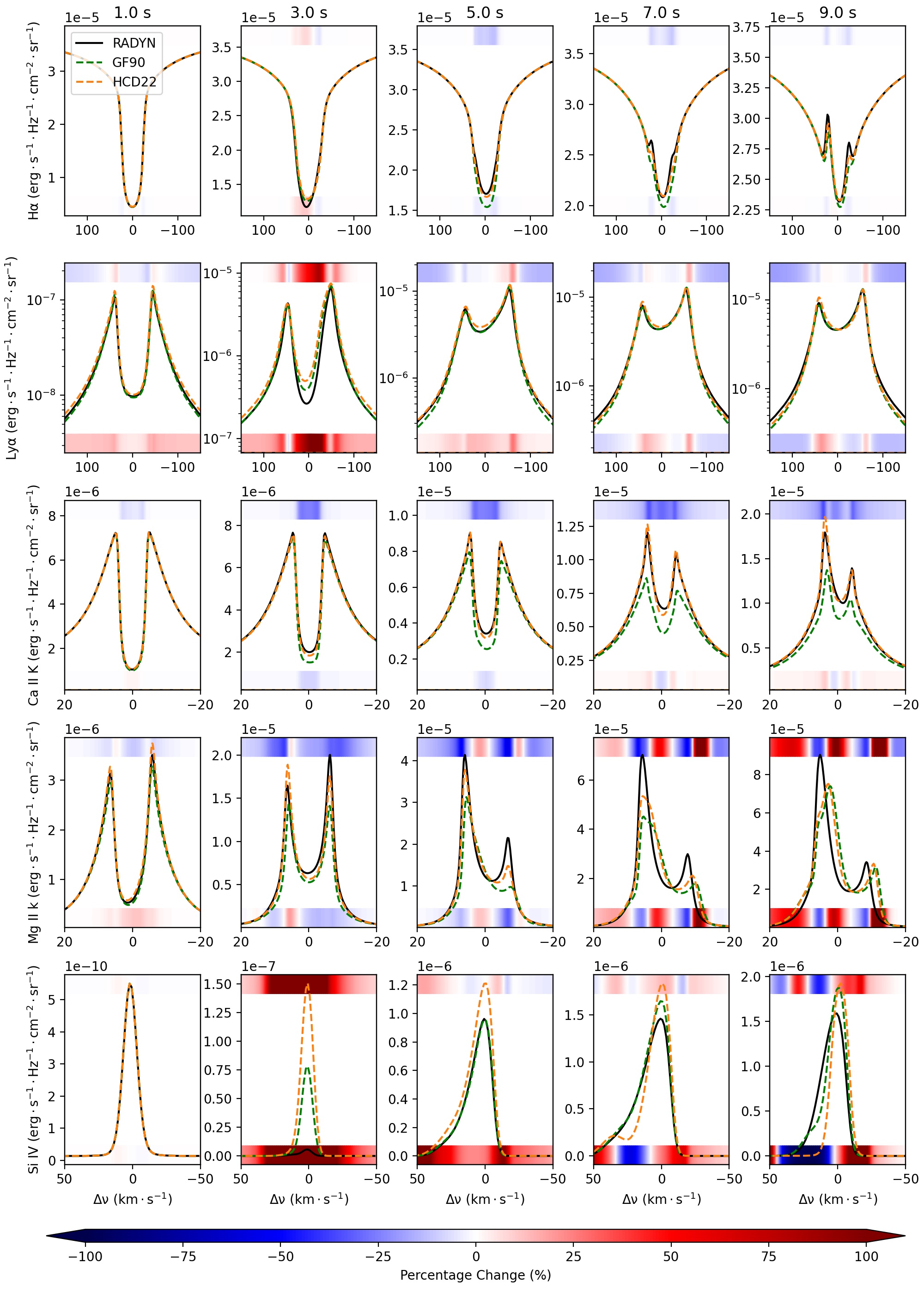}
\caption{Same as Figure \ref{fig:spectram_fa}, but for FC.}
\label{fig:spectram_fc}
\end{figure*}

\begin{figure*}
\includegraphics[width=17cm]{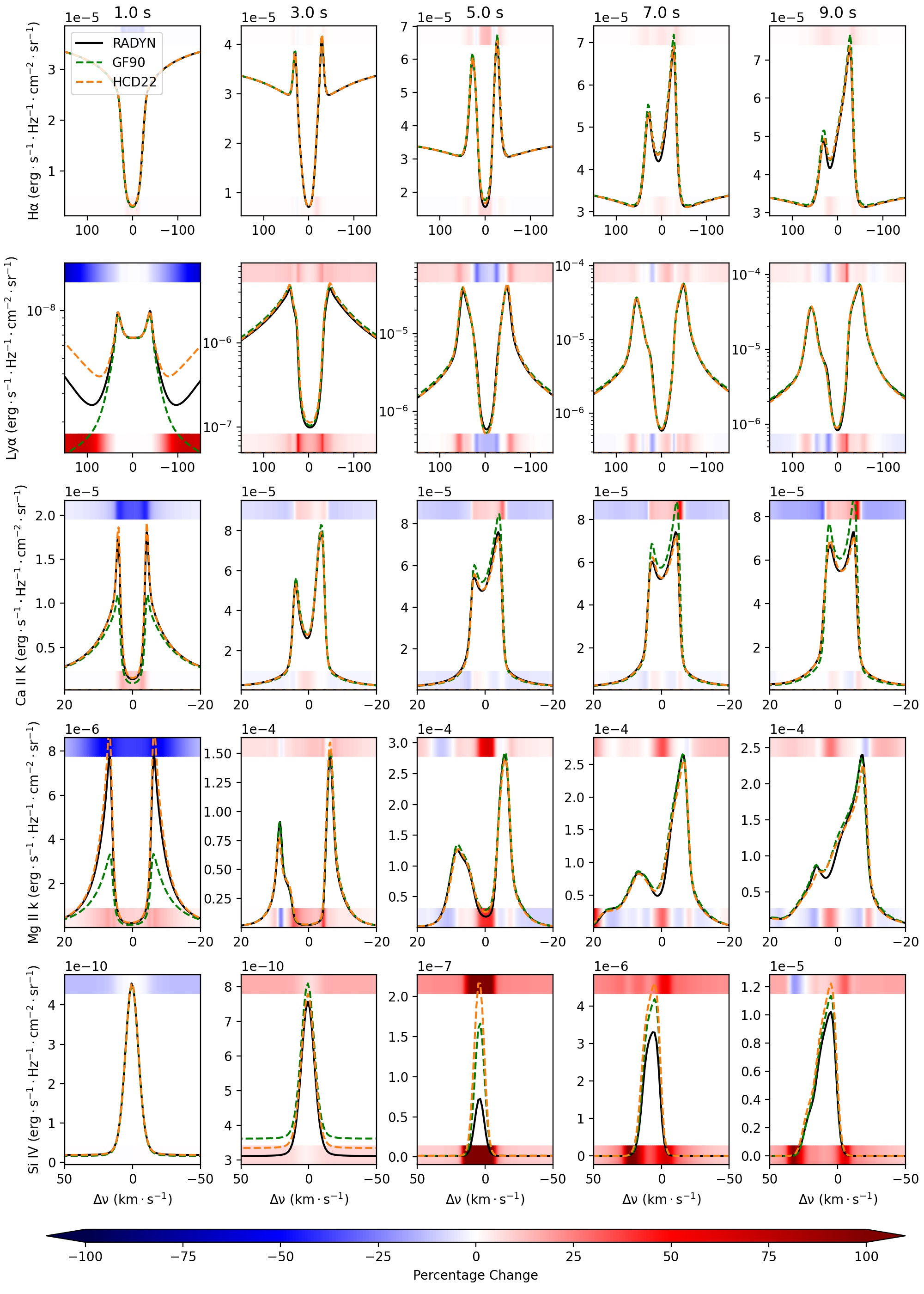}
\caption{Same as Figure \ref{fig:spectram_fa}, but for FD.}
\label{fig:spectram_fd}
\end{figure*}

The spectral line profiles are direct observables that could reflect the atmospheric structure and flare dynamics. We calculate selected line profiles for all the flare models above. Similar to \citet{2022A&A...661A..77H}, the H$\alpha$ line profiles are taken directly from \verb"RADYN", while the Ly$\alpha$, \ion{Ca}{II} K, and \ion{Mg}{II} k line profiles are calculated using the \verb"RH" code \citep{2001ApJ...557..389U,2015A&A...574A...3P}. The \ion{Si}{IV} 1403 \AA\ line profiles are calculated using \verb"MS_RADYN" following \citet{2019ApJ...871...23K}. The results are shown in Figs.~\ref{fig:spectram_fa}--\ref{fig:spectram_fd}, respectively.

Case FA1 has a cooler chromosphere than Case FA0 at 1.0 s, which leads to a weaker Ly$\alpha$ line wing, a weaker H$\alpha$ line center, and weaker \ion{Ca}{II} K$_2$ and \ion{Mg}{II} k$_2$ peaks.  After a certain time of flare heating, the chromosphere of Case FA1 is warmer than that of Case FA0, causing a stronger emission in the H$\alpha$,  \ion{Ca}{II} K and \ion{Mg}{II} k lines, as well as in the Ly$\alpha$ line wings. We notice that the continuum near the \ion{Si}{IV} line is also enhanced at 5.0 s. The Ly$\alpha$ line center and the \ion{Si}{IV} 1403 \AA\ line are relatively weaker due to the lower temperature in the upper chromosphere. Apart from the intensity differences, the line asymmetries in Cases FA0 and FA1 are mostly the same.

\begin{figure*}
\includegraphics[width=14.4cm]{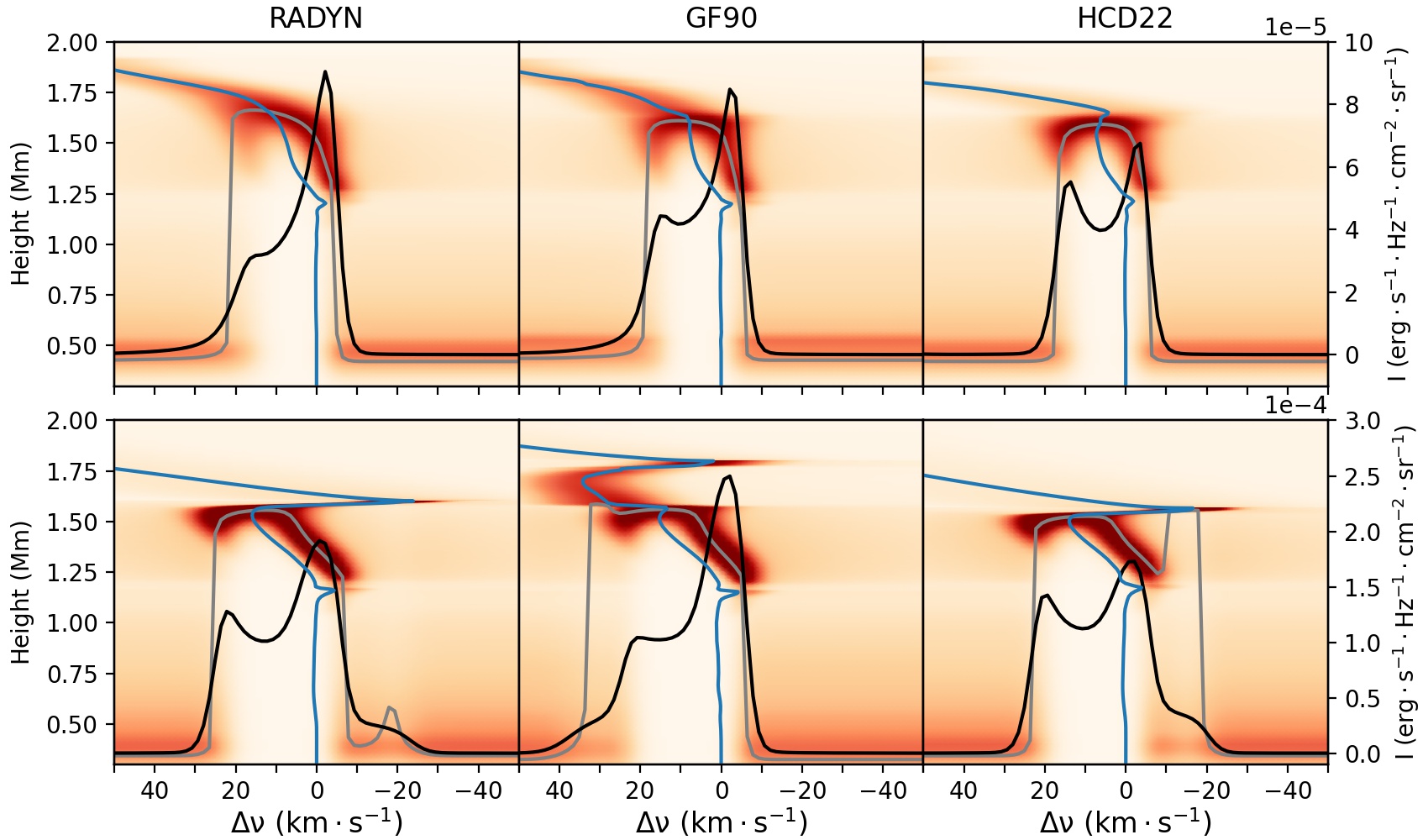}
\caption{The \ion{Si}{IV} line formation at 5.0 s (top row) and 9.0 s (bottom row) calculated under different radiative losses in Case FB. The background is the contribution function to the emergent intensity. The black lines represent the line profile and the grey lines represent the height at which optical depth unity is reached. The blue lines show the vertical velocity.}
\label{fig:fb_si4}
\end{figure*}

Similarly, Case FB1 also underestimates the intensity of H$\alpha$, Ly$\alpha$, \ion{Ca}{II}, and \ion{Mg}{II} lines in the early stage of flare evolution (see the first column in Fig. \ref{fig:spectram_fb}). When heating proceeds, the H$\alpha$, \ion{Ca}{II} and \ion{Mg}{II} lines show  stronger emissions.
The H$\alpha$ line width is also significantly enhanced.
The largest difference lies in the line profiles of Ly$\alpha$ and \ion{Si}{IV} at 9.0 s. 
With the appearance of the condensation region,
the Ly$\alpha$ line center is gradually shifted from blue to red in Case FB0. However, in Case FB1 the Ly$\alpha$ line center is still blueshifted. 
As for the \ion{Si}{IV} line, the opacity is taking effect so that a central reversal gradually appears. The blueshifted line center at 5.0 s corresponds to the  mass flows at the bottom of the evaporating chromospheric plasma as shown in Fig. \ref{fig:fb_si4}.  In Case FB1, the profile shows a smaller blueshift of the line center and hence a weaker asymmetry compared to Case FB0. At 9.0 s, there is a hump in the red wing near $-$20 km s$^{-1}$ in Case FB0, corresponding to the condensation downflow at 1.6 Mm. While in Case FB1, the upflows in the upper chromosphere (1.6--1.7 Mm) contribute to the blue-wing hump near 30 km s$^{-1}$, and the adjacent region (around 1.8 Mm) gives rise to the peak emission near the line core.

For Case FC1, the lower temperature below 1.5 Mm caused by the absence of radiative heating results in the underestimation of the intensities of the H$\alpha$, \ion{Ca}{II} and \ion{Mg}{II} lines almost in the entire evolutionary stage as well as the Ly$\alpha$ line during the later stages of heating. And the \ion{Mg}{II} line shifts to a longer wavelength due to a downflow in the region near 1.53 Mm. 
For Case FD1, the differences in the H$\alpha$, \ion{Ca}{II} and \ion{Mg}{II} line profiles  compared with Case FD0 are similar to the ones between Cases FA1 and FA0. However, unlike Case FA1, FD1 overestimates the whole Ly$\alpha$ line and underestimates the \ion{Ca}{II} line wing for nearly the whole process. For the \ion{Si}{IV} line,  both FC1 and FD1 overestimate the intensity near line core due to the overestimation of the temperature in the upper chromosphere (1.55--1.7 Mm) where the line is formed.
The line asymmetries in FC1 are mostly similar to the ones in FC0, and the blue asymmetry of the \ion{Si}{IV} line at 9.0 s is more pronounced. Case FD1 shows almost the same line asymmetries as Case FD0.

\subsection{Comparison of HCD22 and RADYN}
\subsubsection{Atmospheric structure}
\begin{figure*}
\includegraphics[width=18.4cm]{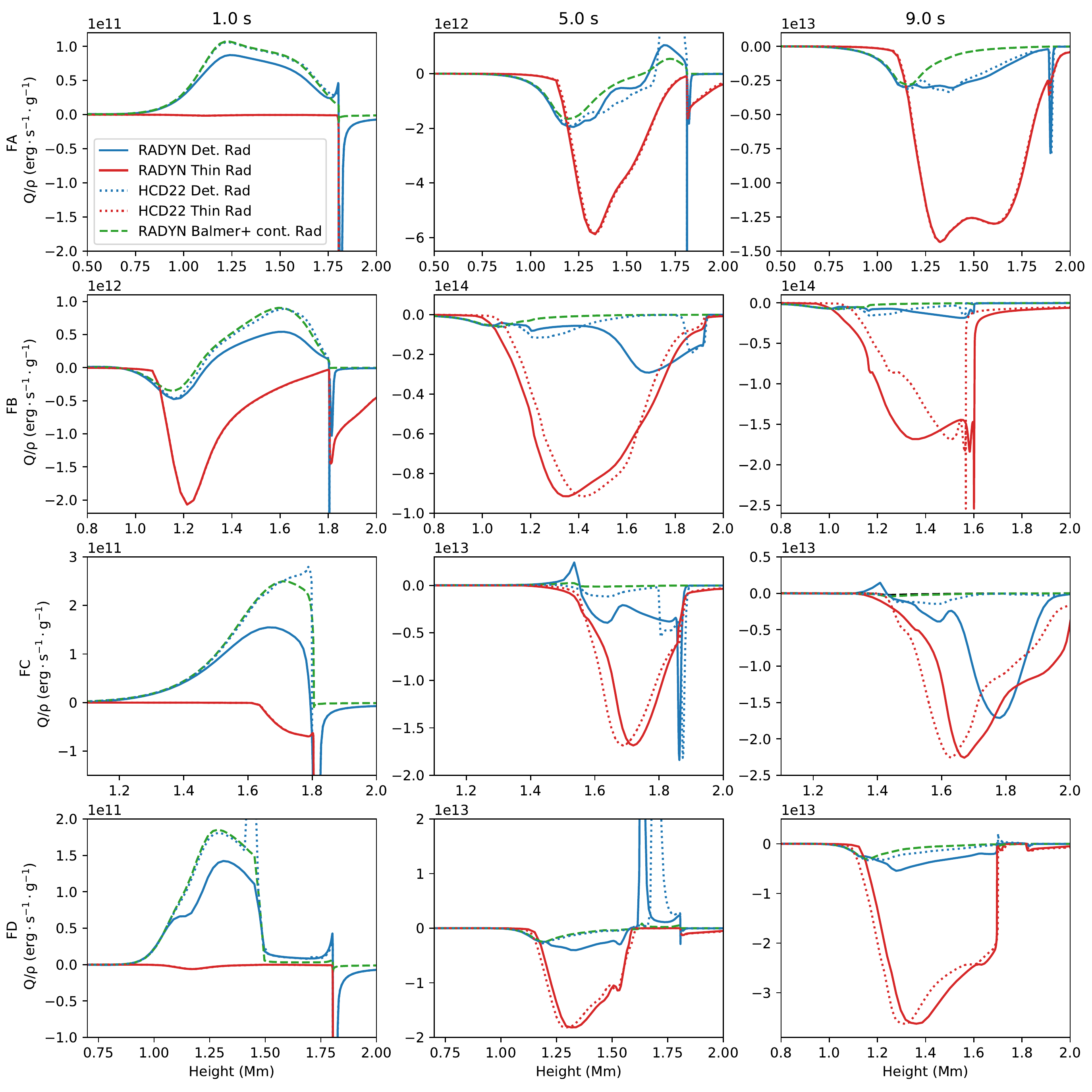}
\caption{Height distributions of the radiative losses per unit mass ($Q/\rho$) calculated with detailed treatment of the radiative processes by RADYN (solid lines) and by the recipe of \citetalias{2022A&A...661A..77H} (dotted lines). From top to bottom, the four rows represent the results for Cases FA, FB, FC and FD. The blue and red lines represent the losses calculated with detailed treatment and with optically thin assumption in each panel respectively. Also plotted is the radiative losses from H Balmer and higher continua (green dashed lines).
A positive value means radiative heating, while a negative one means radiative cooling.}
\label{fig:radloss_hcd22}
\end{figure*}

Fig. \ref{fig:atmosphere} also shows the atmospheric structure calculated from detailed treatment (FA0--FD0) and the approximated recipe of \citetalias{2022A&A...661A..77H} (FA2--FD2) for radiative losses. It is noted that the radiative losses from continua in the FA2--FD2 models are calculated in the same way as in the FA0--FD0 models, instead of using an approximated recipe. Overall, the recipe of \citetalias{2022A&A...661A..77H} gives the accurate estimation of integrated chromospheric radiative losses as shown in Table \ref{tab:tot_loss}. In the initial atmosphere for Case FA2 at 1.0 s, the recipe of \citetalias{2022A&A...661A..77H} underestimates optically thick radiative cooling above 1.1 Mm as shown in Fig. \ref{fig:radloss_hcd22}, and thus the net radiative heating is larger than the one in Case FA0, resulting in a slightly warmer chromosphere. At 5.0 s, an obvious temperature enhancement is seen above 1.6 Mm, which is attributed from the extensive radiative heating from LyC. However, as heating proceeds, radiative heating from LyC rapidly decreases, and the temperature structure of Case FA2 resembles that of Case FA0. The total radiative cooling is dominated by optically thin line cooling.

For Case FB2, the chromospheric temperature above 1.4 Mm rises more rapidly, due to an underestimation of line coolings as a result of an underestimation of the \ion{H}{I} number density. Compared with Case FB0, more intensive chromospheric evaporation drives more plasma moving upwards into the corona, and the chromospheric condensation region appears earlier in time and with larger downflow velocities as shown in Fig. \ref{fig:velocity}.
In addition, Case FB2 has a  cooler  chromosphere at the height range of  1.0--1.3 Mm clearly in the later stage due to a larger cooling in this region.

The upper chromosphere in Cases FC2 and FD2 behaves similarly to that in Case FB2. However, similar to Case FC1, we also notice a cool dense region in Case FC2 at around 1.4 Mm which results from the absence of heating from LyC. A larger downflow of about 6 km s$^{-1}$ also exists near 1.53 Mm in Case FC2. 

\subsubsection{Line profiles}
The tiny overestimation of temperature for Case FA2 in the middle chromosphere leads to a stronger emission of the Ly$\alpha$ line wing, and \ion{Ca}{II} K$_2$ and \ion{Mg}{II} k$_2$ peaks at 1.0 s. At 5.0 s, the warmer chromosphere above 1.5 Mm in Case FA2 contributes to the enhancements at the line centers of the H$\alpha$, Ly$\alpha$ and \ion{Si}{IV} lines. Generally speaking, Case FA2 produces fairly consistent results to Case FA0, in terms of line intensity and asymmetries.

Because of the earlier appearance of the condensation region in Case FB2, the change of asymmetry in the Ly$\mathrm{\alpha}$ line from red to blue occurs earlier at 7.0 s in Case FB2, as well as the red-wing hump in the \ion{Si}{IV} line. The blueshift of \ion{Si}{IV} line center is smaller at 5.0 s as a result of the relatively moderate upflow as shown in Fig. \ref{fig:fb_si4}. 
In addition, the cooler chromosphere in the height range of 1.0--1.3 Mm  results in a weaker intensity near the \ion{Ca}{II} and \ion{Mg}{II} line centers in the later stage. 

In Case FC2, the overestimation of temperature in the upper chromosphere results in a stronger emission near the Ly$\mathrm{\alpha}$ and \ion{Si}{IV} line centers. However, the blue asymmetry of the \ion{Si}{IV} line is less pronounced, because the upflow region is intensively heated so that the temperature has exceeded the formation temperature of the \ion{Si}{IV} line. The downflow near 1.53 Mm produces  a more redshifted \ion{Mg}{II} line center. 

In Case FD2, strong enhancements in the Ly$\mathrm{\alpha}$ line wings and the \ion{Ca}{II} and \ion{Mg}{II} emission peaks appear at 1.0 s, similar to Case FA2. The differences in these chromospheric lines become less obvious as heating proceeds. However, the \ion{Si}{IV} line emission is enhanced greatly in Case FD2, similar to Case FD1. The line asymmetries in Case FD2 are mostly similar to those in Case FD0.

\section{Discussion \label{discussion}}
\subsection{Line transitions in RADYN \label{4.1}}
In \verb"RADYN", all the line transitions are calculated based on the assumption of complete frequency redistribution (CRD)  \citep{1992ApJ...397L..59C}. The Lyman series are, however, treated with a Gaussian line profile in order to mimic the partial frequency redistribution (PRD) effects \citep{2012ApJ...749..136L}. This turns out to be a good estimation, since the Ly$\alpha$ line profiles are quite close to the ones in PRD \citep{2019ApJ...879..128H}. For the \ion{Ca}{II} and \ion{Mg}{II} lines with strong coherent scattering, the radiative losses under CRD will be overestimated since the line-center photons can be scattered to the line wing and then escape freely \citep{2002ApJ...565.1312U,1995A&A...293..166H}. Thus, in practice the \ion{Mg}{II} lines are neglected in \verb"RADYN", in order to prevent such an overestimation of the radiative losses.

The radiative losses from H in the \citetalias{2022A&A...661A..77H} recipe are based on detailed calculations from \verb"RADYN", while the losses from Ca and Mg are based on the results from \verb|RH| under the assumption of PRD and statistical equilibrium (SE). Note the difference in the codes \verb|RH| and \verb"RADYN" when treating the lines and level populations, i.e., PRD vs. CRD, as well as SE vs. non-equilibrium ionization (NE).  We list the total radiative losses from the H, Ca, and He atoms in \verb"RADYN", and from the Ca and Mg atoms in \verb"RH", for the four flare models (FA0--FD0) in Table \ref{tab:flare_flux}. In the flare chromosphere, the total integrated radiative loss from \ion{Mg}{II} h \& k and \ion{Ca}{II} H \& K is more  significant than that from the \ion{Ca}{II} IR triplet, which is consistent with the results corresponding to the footpoint of a flare loop in \citet{2022arXiv220702840Y}. While the contribution from the Ly$\alpha$ is underestimated significantly in \citet{2022arXiv220702840Y} due to their treatment of transition region. \citet{2019ApJ...885..119K} argued that SE only affects line intensities in the initial heating and cooling phases. However, we find that in our models, SE can take effect during the whole heating phase. That is why the radiative losses from \verb"RADYN" (NE+CRD) are still smaller than those from \verb|RH| (SE+PRD) in Case FB0 at 9.0s. This discrepancy is attributed to the different heating functions in the flare models, which is constant in \citet{2019ApJ...885..119K} while a linearly increasing function in our models.
One can also see that the total radiative losses from Ca and Mg in \verb|RH| are far larger than those from Ca in \verb"RADYN". Therefore, the losses from Ca and Mg in the middle chromosphere (around 1.2--1.4 Mm) in the FA2--FD2 models exceed those in the FA0--FD0 models. In fact, the neglection of Mg in \verb"RADYN" might not be an accurate enough approximation, neither is the SE assumption in the recipes. A self-consistent inclusion of Ca and Mg under both NE and PRD assumptions in the flare models is currently unavailable and deserves investigations in the future. 

Radiative losses from He also have a significant proportion as suggested in Table \ref{tab:flare_flux}, which are dominant especially in the upper chromosphere (1.5--1.8 Mm). However, the largest contributions result from the \ion{He}{I} 584 \AA\ and \ion{He}{II} 304 \AA\ lines, which are usually considered optically thin \citep{2017A&A...597A.102G}. Thus, in the FB2 and FC2 cases where losses from He are absent, we do see an underestimation of the radiative losses in the upper chromosphere. In practice, this could be implemented by including the optically thin losses from the EUV lines of He. The optically thick \ion{He}{I} 10830 \AA\ line is, however, another important candidate that needs proper consideration. Radiative losses from this line amounts to about 26\% of the total value from He in the chromosphere for Case FA0.

\begin{table*}[h]
    \caption{Temporally and spatially integrated radiative losses (in units of 10$^9$ $\mathrm{erg \cdot cm ^{-2}}$) and spatially integrated radiative losses (in units of 10$^{8}$ $\mathrm{erg \cdot cm ^{-2} \cdot s^{-1}}$) at 9.0 s of chromosphere in the height range of 0.5-–1.8 Mm from different atoms and lines in different models.}
    \centering 
    \tabcolsep=0.1cm
    \begin{tabular}{lcccccccccccc} 
        \hline\hline
        \multicolumn{1}{l}{Flare} & \multicolumn{2}{c}{H (RADYN)} & \multicolumn{1}{c}{\ion{Ca}{II} (RADYN)} & \multicolumn{1}{c}{He (RADYN)} & \multicolumn{4}{c}{\ion{Ca}{II} (RH)} & & \multicolumn{3}{c}{\ion{Mg}{II} (RH)}  \\
        \cline{2-3}
        \cline{6-9}
        \cline{11-13}
        & Total & Ly$\alpha$ & Total & Total & Total & \ion{Ca}{II} H & \ion{Ca}{II} K & \ion{Ca}{II} IR & & Total & \ion{Mg}{II} h & \ion{Mg}{II} k \\
        \hline
        \multicolumn{13}{c}{Temporally and spatially integrated radiative losses} \\
        \hline
        FA0 &  1.51 & 0.17 &         0.32&  0.25 &         0.19 & 0.05 & 0.06 & 0.09 & &        0.26 & 0.11 & 0.14 \\
        FB0 &  7.99 & 1.31 &         0.50&  6.69 &         0.57 & 0.12 & 0.29 & 0.16 &  &       0.59 & 0.26 & 0.32 \\
        FC0 &  0.21 & 0.06 &         0.09&  0.30 &         0.08 & 0.01 & 0.04 & 0.03 &   &      0.09 & 0.04 & 0.06 \\
        FD0 &  2.54 & 0.50 &         0.24&  0.29 &         0.15 & 0.04 & 0.04 & 0.07 &    &     0.19 & 0.09 & 0.10 \\
        \hline
        \multicolumn{13}{c}{Spatially integrated radiative losses at 9.0 s} \\
        \hline
        FA0 &  3.67 & 0.64 &         0.47&  0.46 &         0.29 & 0.08 & 0.09 & 0.12 &  &         0.43 & 0.19 & 0.25 \\
        FB0 &  8.15 & -0.28 &         0.68&  13.88 &         0.86 & 0.19 & 0.45 & 0.22 & &         0.94 & 0.42 & 0.52 \\ 
        FC0 &  0.54 & 0.22 &         0.11&  0.71 &         0.10 & 0.01 & 0.05 & 0.04  & &         0.18 & 0.07 & 0.10 \\
        FD0 &  5.17 & 1.11 &         0.21&  0.50 &         0.20 & 0.05 & 0.06 & 0.08  &  &       0.34 & 0.15 & 0.19 \\
        \hline
    \end{tabular}
    \label{tab:flare_flux}
\end{table*}

\subsection{Radiative heating from LyC \label{4.2}}
During the flare, radiative heating also plays an important role in the thermodynamic evolution of the atmosphere \citep{1980ApJ...242..336M,2012A&A...539A..39C}.  The main sources of radiative heating in solar flares include LyC as well as the Ly$\alpha$ line \citep{1980ApJ...242..336M,2019ApJ...882...97P,2022A&A...661A..77H}. The H Balmer and higher continua also act as heating sources in the first few seconds during the flare evolution, and they switch to strong cooling sources once the atmosphere is intensively heated.  The absence of radiative heating in the recipes would naturally lead to a cooler atmosphere, as in the first few seconds in Cases FA1--FD1.

\begin{figure*}
\includegraphics[width=14.4cm]{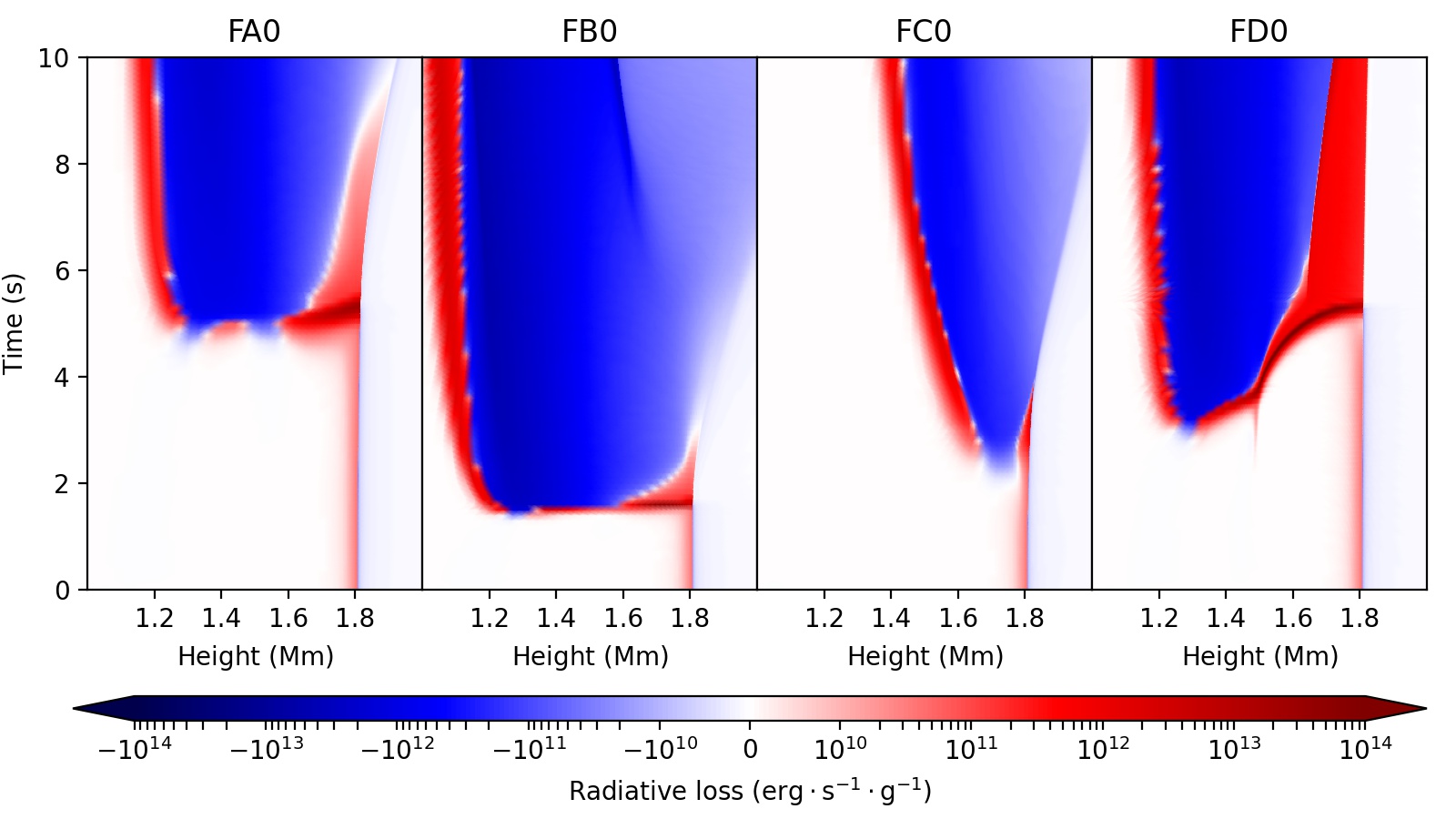}
\caption{Radiative loss from LyC in the FA0--FD0 cases. A positive value means radiative heating, while a negative one means radiative cooling.}
\label{fig:rad_lyc}
\end{figure*}

Different from the quiet Sun where Ly$\alpha$ is the main radiative heating source \citep{2012A&A...539A..39C}, in flare conditions LyC dominates in radiative heating \citep{2022A&A...661A..77H}. Radiative losses from LyC are sketched in Fig. \ref{fig:rad_lyc}  for the models FA0--FD0.  One can see that in the middle chromosphere, LyC contributes to radiative cooling, while there appears to be two regions where LyC contributes to radiative heating, one below and one above the cooling region. The region below the cooling region is heated through the absorption of downward propagating LyC photons, usually referred to as ``backwarming''. The backwarming region, however, suffers a strong cooling from the Balmer continuum. Thus, the energy from backwarming is effectively radiated away so that the temperature is not enhanced much, and the backwarming process can last through the whole flare heating process. 
For the region above the cooling region, the upward propagating LyC photons from the middle chromosphere are absorbed, since this region is not heated much and remains to have a relatively lower temperature. Radiative cooling in this region is very weak due to a relatively lower density. Therefore, the LyC photons could heat the local atmosphere effectively, and one can see a quick rise in the temperature. After that, radiative heating gradually ceases due to the drop of the number density of \ion{H}{I}. As a result, the heating process appears to be very impulsive and lasts only for a short time. This process seems to be quite important in flare dynamics, and further investigations are required. For the moment, we choose to keep radiative heating in this region in our simulations.

\subsection{Performance of the recipes}
As stated above, \citetalias{1990ApJ...358..328G} overestimates the radiative cooling in the upper chromosphere, resulting in a cooler atmosphere with a weaker chromospheric evaporation. Especially, when the temperature exceeds 10$^5$ K, the overestimation of radiative losses is enormous, which nearly halts the rise of temperature and the appearance of a condensation region. The cooling in the middle and lower chromosphere, however, is underestimated. The difference in  temperature leads to a misestimation of the line intensities, especially for those formed in the chromosphere. The line asymmetry of Ly$\alpha$ and \ion{Si}{IV} will be misestimated due to the absence of condensation region. For the recipe of \citetalias{2022A&A...661A..77H}, the performances are more or less acceptable, though there are still underestimations of cooling due to the underestimation of \ion{H}{I} density in the upper chromosphere. The line intensities also deviate a little from, but the asymmetry is consistent with the results from detailed calculations by \verb"RADYN". 

Therefore, if one intends to analyze the chromospheric dynamics and focus on the spectral lines during a flare, especially those lines that are formed in the chromosphere, the recipe of \citetalias{2022A&A...661A..77H} would be a better choice. However, the total cooling from the recipes of \citetalias{1990ApJ...358..328G} and \citetalias{2022A&A...661A..77H} is similar, and they both provide a reasonable evaluation of the chromospheric cooling. Both recipes are good choices if one only needs to estimate the radiative loss of the entire chromosphere without considering the detail spatial distribution. Technically, the recipe of \citetalias{1990ApJ...358..328G} would be more feasible since only the local variables (temperature and density) are needed to calculate the radiative losses, while the recipe of \citetalias{2022A&A...661A..77H} takes the column density as an input, which is an integrated variable.

Moreover, if one intends to calculate the radiative losses in observations, the method proposed by \citet{2022arXiv220702840Y} would be a good choice. This approach could provide an observational constricted map of radiative losses using the \verb"STiC" inversion code  \citep{2016ApJ...830L..30D,2019A&A...623A..74D}.
\section{Conclusion \label{conclusion}}
In this paper, we evaluate the performances of two recipes for calculating chromospheric radiative losses in flare conditions. We find that both recipes give similar results of the total chromospheric cooling, which are tested to be a good approximation to the real value. For weak flares, both recipes could generate line profiles of similar shapes, although the intensity could be different to some extent. For strong flares, especially when the chromosphere is heated to more than 10$^5$ K, the recipe of \citetalias{1990ApJ...358..328G}  overestimates  cooling in the upper chromosphere, which hinders the appearance of a condensation region. The recipe of \citetalias{2022A&A...661A..77H} underestimates cooling, and thus speeds up the chromospheric evaporation and condensation processes. Therefore, we suggest switching to \citetalias{2022A&A...661A..77H} when simulating strong flares with the peak electron flux exceeding the 10$^{11}$ $\mathrm{erg \cdot cm^{-2} \cdot s^{-1}}$ threshold. While for weak flares with the peak electron flux lower than  the threshold, the recipe of \citetalias{1990ApJ...358..328G} seems to be a better choice due to its simple form. Realistic or test magnetohydrodynamic simulations of flares with these recipes are thus desirable in order to further check their applicability.

Radiative heating is a long-standing issue that neither recipe could take into consideration. From our results, it seems that radiative heating in the middle and lower chromosphere could be safely neglected, if one is only interested in the impulsive phase of a flare. Radiative heating in these regions only dominates in the first few seconds, and only influences the line intensity, with no signs of plasma flows as revealed from line asymmetries. However, radiative heating from LyC in the upper chromosphere seems to be quite important, and a detailed investigation would be required.

\begin{acknowledgements}
      We would like to thank the referee for the comments and suggestions. This work was supported by National Key R\&D Program of China under grant 2021YFA1600504 and by NSFC under grant 11903020, 11873095, 11733003, and 12127901. Y.L. is also supported by the CAS Pioneer Talents Program for Young Scientists and by the CAS Strategic Pioneer Program on Space Science under grants XDA15052200, XDA15320103, and XDA15320301. We thank Dr. Graham S. Kerr for providing the Si model atom. We appreciate the open-source packages: SciPy, Matplotlib and NumPy.
\end{acknowledgements}


%
   \bibliographystyle{aa} 
   \bibliography{aa.bib} 
%

\begin{appendix}
\section{Comparison of the evolution of the electron density}

\begin{figure*}
\includegraphics[width=18.4cm]{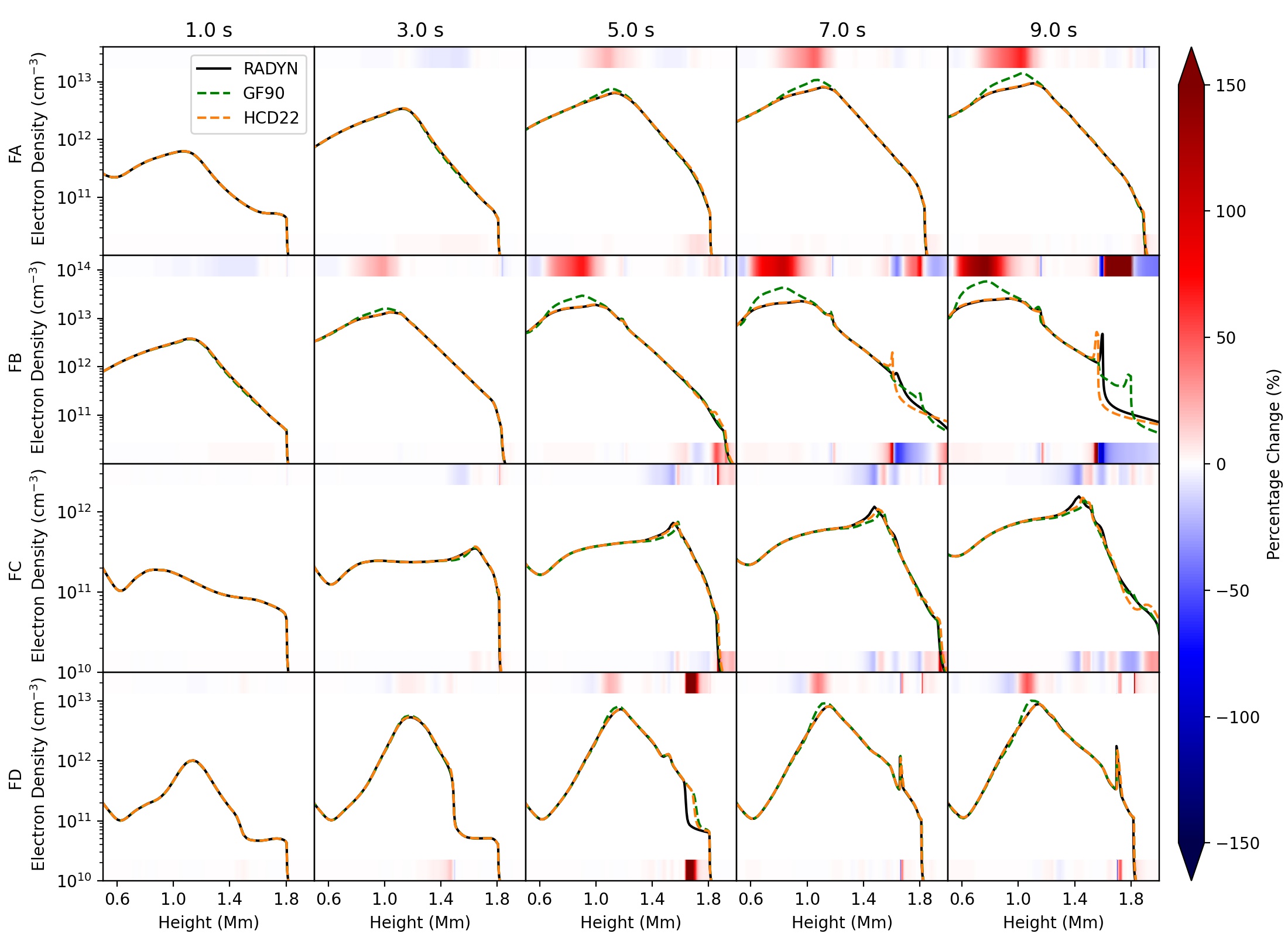}
\caption{Comparison of the evolution of the electron density. The top row shows the density calculated with detailed treatment of the radiative processes by RADYN (black solid lines) as well as that calculated using the recipe of \citetalias{1990ApJ...358..328G} (green dashed lines) and \citetalias{2022A&A...661A..77H} (orange dashed lines) for Case FA. The following three rows are the same as the top row, but for Cases FB, FC and FD, respectively. The density deviation between \citetalias{1990ApJ...358..328G} and RADYN results is also shown as a horizontal bar at the top of each panel, and that between \citetalias{2022A&A...661A..77H} and RADYN results is shown at the bottom of each panel in a color scale.}
\label{fig:density}
\end{figure*}

\end{appendix}

\end{document}